\newif\iffigs\figstrue

\documentstyle[12pt]{article}
\iffigs
  \input epsf
\else
\fi

\batchmode
  \newfont{\footscrfont}{rsfs10}
  \newfont{\footbbbfont}{msbm10}
\errorstopmode

\newif\ifscrf\scrftrue
\ifx\footscrfont\nullfont
  \scrffalse
\fi

\newif\ifamsf\amsftrue
\ifx\footbbbfont\nullfont
  \amsffalse
\fi

\def\ppnumber{\vbox{\baselineskip14pt\hbox{CLNS-96/1402}
\hbox{hep-th/9602118}}}
\def\ppdate{February 1996}
\def\pplogo{\vbox{\kern-\headheight\kern -15pt
\halign{##&##\hfil\cr&{
\ppnumber}\cr\rule{0pt}{2.5ex}&\ppdate\cr}
}}

\makeatletter
\date{}
\def\dedicatory#1{\def\@date{\normalsize\it#1}}
\def\subjclass#1{\def\@thefnmark{}\@footnotetext{1991
    {\it Mathematics Subject Classification.} #1}}
\def\keywords#1{\def\@thefnmark{}\@footnotetext{
    {\it Key words and phrases.} #1}}

\def\ps@firstpage{\ps@empty \def\@oddhead{\hss\pplogo}%
  \let\@evenhead\@oddhead 
}
\def\maketitle{\par
 \begingroup
 \def\thefootnote{\fnsymbol{footnote}}
 \def\@makefnmark{\hbox
 to 0pt{$^{\@thefnmark}$\hss}}
 \if@twocolumn
 \twocolumn[\@maketitle]
 \else \newpage
 \global\@topnum\z@ \@maketitle \fi\thispagestyle{firstpage}\@thanks
 \endgroup
 \setcounter{footnote}{0}
 \let\maketitle\relax
 \let\@maketitle\relax
 \gdef\@thanks{}\gdef\@author{}\gdef\@title{}\let\thanks\relax}

\def\abstract{\if@twocolumn
\section*{Abstract}
\else \small
\begin{center}
{\bf ABSTRACT}
\end{center}
\quotation
\fi}

\def\thebibliography#1{\section*{References\@mkboth
 {REFERENCES}{REFERENCES}}\small\list
 {[\arabic{enumi}]}{\settowidth\labelwidth{[#1]}\leftmargin\labelwidth
 \advance\leftmargin\labelsep
 \usecounter{enumi}}
 \def\newblock{\hskip .11em plus .33em minus .07em}
 \sloppy\clubpenalty4000\widowpenalty4000
 \sfcode`\.=1000\relax}

\newif\iffn\fnfalse

\@ifundefined{reset@font}{\let\reset@font\empty}{} 
\long\def\@footnotetext#1{\insert\footins{\reset@font\footnotesize
    \interlinepenalty\interfootnotelinepenalty
    \splittopskip\footnotesep
    \splitmaxdepth \dp\strutbox \floatingpenalty \@MM
    \hsize\columnwidth \@parboxrestore
   \edef\@currentlabel{\csname p@footnote\endcsname\@thefnmark}\@makefntext
    {\rule{\z@}{\footnotesep}\ignorespaces
      \fntrue#1\fnfalse\strut}}}

\newcount\@tempcntc
\def\@citex[#1]#2{\if@filesw\immediate\write\@auxout{\string\citation{#2}}\fi
  \@tempcnta\z@\@tempcntb\m@ne\def\@citea{}\@cite{\@for\@citeb:=#2\do
    {\@ifundefined
       {b@\@citeb}{\@citeo\@tempcntb\m@ne\@citea
        \def\@citea{,\penalty\@m\ }{\bf ?}\@warning
       {Citation `\@citeb' on page \thepage \space undefined}}%
    {\setbox\z@\hbox{\global\@tempcntc0\csname b@\@citeb\endcsname\relax}%
     \ifnum\@tempcntc=\z@ \@citeo\@tempcntb\m@ne
       \@citea\def\@citea{,\penalty\@m}
       \hbox{\csname b@\@citeb\endcsname}%
     \else
      \advance\@tempcntb\@ne
      \ifnum\@tempcntb=\@tempcntc
      \else\advance\@tempcntb\m@ne\@citeo
      \@tempcnta\@tempcntc\@tempcntb\@tempcntc\fi\fi}}\@citeo}{#1}}

\def\@citeo{\ifnum\@tempcnta>\@tempcntb\else\@citea
  \def\@citea{,\penalty\@m}%
  \ifnum\@tempcnta=\@tempcntb\the\@tempcnta\else
   {\advance\@tempcnta\@ne\ifnum\@tempcnta=\@tempcntb \else
\def\@citea{--}\fi
    \advance\@tempcnta\m@ne\the\@tempcnta\@citea\the\@tempcntb}\fi\fi}

\makeatother

\ifamsf
  \newfont{\bigbbbfont}{msbm10 scaled\magstep2}
  \newfont{\bbbfont}{msbm10 scaled\magstep1}  
  \newfont{\smallbbbfont}{msbm8}
  \newfont{\tinybbbfont}{msbm6}
  \newfont{\smallfootbbbfont}{msbm7}
  \newfont{\tinyfootbbbfont}{msbm5}
\fi

\ifscrf
  \newfont{\scrfont}{rsfs10 scaled\magstep1}  
  \newfont{\smallscrfont}{rsfs7}
  \newfont{\tinyscrfont}{rsfs7}
  \newfont{\smallfootscrfont}{rsfs7}
  \newfont{\tinyfootscrfont}{rsfs7}
\fi

\ifamsf
  \newcommand{\Bbb}[1]{\iffn
      \mathchoice{\mbox{\footbbbfont #1}}{\mbox{\footbbbfont #1}}
      {\mbox{\smallfootbbbfont #1}}{\mbox{\tinyfootbbbfont #1}}\else
      \mathchoice{\mbox{\bbbfont #1}}{\mbox{\bbbfont #1}}
      {\mbox{\smallbbbfont #1}}{\mbox{\tinybbbfont #1}}\fi}
\else
  \def\bigbbbfont{\bf}
  \def\Bbb{\bf}
\fi

\ifscrf
  \newcommand{\Scr}[1]{\iffn
    \mathchoice{\mbox{\footscrfont #1}}{\mbox{\footscrfont #1}}
    {\mbox{\smallfootscrfont #1}}{\mbox{\tinyfootscrfont #1}}\else
    \mathchoice{\mbox{\scrfont #1}}{\mbox{\scrfont #1}}
    {\mbox{\smallscrfont #1}}{\mbox{\tinyscrfont #1}}\fi}
\else
  \def\Scr{\cal}
\fi

\def\P{{\Bbb P}}

\def\Z{{\Bbb Z}}

\def\opeq#1{\advance\lineskip#1 \advance\baselineskip#1
        \advance\lineskiplimit#1}

\def\eqalign#1{\null\,\vcenter{\opeq{2.5\jot}\mathsurround=0pt
        \everycr={}\tabskip=0pt
        \halign{\strut\hfil$\displaystyle{##}$&$\displaystyle{{}##}$\hfil
        \crcr#1\crcr}}\,\null}

\def\sm{$\sigma$-model}

\def\CY{Calabi--Yau}

\def\cM{{\Scr M}}

\def\cD{{\Scr D}}

\def\cMc{{\hfuzz=100cm\hbox to 0pt{$\;\overline{\phantom{X}}$}\cM}}
\def\barcD{{\hfuzz=100cm\hbox to 0pt{$\;\overline{\phantom{X}}$}\cD}}

\ifamsf

\else

\fi

\begin{document}
\setcounter{page}0
\title{\LARGE Heterotic-Heterotic String Duality\\ and Multiple
K3 Fibrations\\[10mm]}
\author{
Paul S. Aspinwall\\[0.7cm]
\normalsize F.R.~Newman Lab.~of Nuclear Studies,\\
\normalsize Cornell University,\\
\normalsize Ithaca, NY 14853\\[10mm]
Mark Gross\\[0.7cm]
\normalsize Department of Mathematics,\\
\normalsize Cornell University,\\
\normalsize Ithaca, NY 14853\\[5mm]
}

{\hfuzz=10cm\maketitle}

\def\Large{\large}
\def\LARGE{\large\bf}

\vskip 1cm

\begin{abstract}

A type IIA string compactified on a Calabi--Yau manifold which admits
a K3 fibration is believed to be equivalent to a heterotic string in
four dimensions. We study cases where a Calabi--Yau manifold can have
more than one such fibration leading to equivalences between
perturbatively inequivalent heterotic strings. This allows an analysis
of an example in six dimensions due to Duff, Minasian and Witten and
enables us to go some way to prove a conjecture by Kachru and
Vafa. The interplay between gauge groups which arise perturbatively
and nonperturbatively is seen clearly in this example. As an
extreme case we discuss a Calabi--Yau manifold which admits an
infinite number of K3 fibrations leading to infinite set of equivalent
heterotic strings.

\end{abstract}

\vfil\break


\section{Introduction}

It is now generally believed that the type IIA string compactified
down to six dimensions on a K3 surface gives the same physics as a
heterotic string compactified on a four-torus (see, for example,
\cite{HT:unity,W:dyn}). It was suggested in \cite{KV:N=2,FHSV:N=2}
that much the same effect could be seen in four dimensions for $N=2$
theories. That is, a type
II string compactified on a \CY\ threefold could be equivalent to a
heterotic string compactified on an object, essentially a K3 surface
times a two-torus, for a suitable \CY\ manifold.

Let us suppose that we have a type IIA string compactified on a \CY\
manifold, $X$.
It was realized in \cite{KLM:K3f} that some of the evidence supporting
the conjectured dual pairs of \cite{KV:N=2} appeared to originate in
the fact that $X$ could be written in the form of a K3 fibration. That
is $X$ can be written as a bundle where the generic fibre is a K3
surface and the base space is $\P^1$. Over a finite number of points
on the base $\P^1$ the fibre degenerates and will not be a K3 surface.
It was realized in \cite{VW:pairs} that this fibration structure fits
in nicely with the six-dimensional duality above.

A general result concerning K3 fibrations was derived in
\cite{AL:ubiq}. That is, if there is a heterotic string dual to the
type IIA string compactified on $X$, such that the weakly-coupled
heterotic string can be understood in terms of $X$ near its
large-radius limit, then $X$ {\it must\/} be a K3 fibration. In
particular, the divisor class of the K3 fibre (or equivalently, the
size of the base $\P^1$) on the type IIA side maps to the dilaton on
the heterotic side.

It is natural to ask whether the converse to this statement is
true. That is, if $X$ is a K3 fibration then must there exist a
heterotic dual? Certainly there is a field in the theory which has
every right to be called a dilaton for the heterotic string. This
question is analogous to one
familiar from the relationship between conformal field theories and
non-linear \sm s with a \CY\ target space. When does a conformal field
theory have a \CY\ target space interpretation? A necessary condition
for this is that the moduli space should have a large-radius limit
point, that is a point of {\it maximal unipotency\/}
\cite{Mor:gid}. It seems reasonable to say that this is also a
sufficient condition --- certainly no counter-example is known. This
condition of maximal unipotency appears to be on the same footing as
the K3 fibration condition above and so again it seems very reasonable
to assert that it is again a sufficient condition.

This raises an immediate question. Suppose $X$ is such that it admits
more than one K3 fibration. That is, I can fibre $X$ in two (or more)
ways such that the resulting K3 fibres in each case are not
homologous. We now have two (or more) heterotic strings, each of which can
equally be called the dual of the type IIA theory. Since the K3 fibres
are not homologous, the dilatons for the heterotic strings must
correspond to different directions in the moduli space of vector
moduli. Thus the relationship between the heterotic strings will not
be manifest from the perturbative analysis of the heterotic string
using conformal field theory. Thus, \CY\ manifolds with multiple
K3 fibrations give examples of nontrivial heterotic dual pairs in four
dimensions.

An example of dual heterotic strings in six dimensions was given
recently by Duff, Minasian and Witten \cite{DMW:hh}. Each member of
this pair gives an $N=1$ theory in six
dimensions and so can be compactified on a two-torus to give $N=2$ in
four dimensions. The resulting theory is precisely one discussed in
\cite{KV:N=2}. Thus, if the conjectures of \cite{KV:N=2} are to be
correct then the $X$ given as the dual partner {\it must\/} admit at
least two fibrations. We will see that this is correct, aside from
some minor subtleties, in section \ref{s:eg1}. We will also see this
double fibration structure essentially requires $X$ to be unique.

An important part of the analysis of \cite{DMW:hh} was the necessity
of the possible appearance of gauge groups which could not be
understood perturbatively from the heterotic string. Such groups were
discussed in the context of type IIA strings in \cite{AL:ubiq}. We
will see in section \ref{s:enhg} that the dual fibration picture
discussed in section \ref{s:eg1} fits in perfectly with this
picture. We will also see that the type IIA string predicts more than
can be seen from simple analysis of either of the dual heterotic
strings.

There is actually no limit to how many K3 fibrations some \CY\
manifolds may admit. In section \ref{s:inf} we discuss an extreme case
of an infinite number.

While preparing this paper we became aware of \cite{MV:F} which has
some overlap with this work.


\section{A Double Fibration}   \label{s:eg1}

In this section we will discuss the example of \cite{DMW:hh,KV:N=2} in
the context of a double fibration.

Whereas a type II compactification is specified by the space on which
it is compactified, a heterotic string requires, in addition, a
vector bundle structure in which the gauge fields of the heterotic
string live. The heterotic strings in six dimensions studied in
\cite{DMW:hh} were compactified on a K3 surface with the following
vector bundle. Let $E$ be a stable $SU(2)$-bundle on a K3 with
$c_2(E)=12$. Embed this instanton in an $E_8$-bundle. The $E_8\times
E_8$ bundle required is simply the sum of two such bundles.

Since the structure group of this bundle commutes with an $E_7\times
E_7$ subgroup, we expect to see an $E_7\times E_7$ gauge group in the
resultant six-dimensional physics. This bundle may be deformed
however, that is, there are massless scalars in the six-dimensional
theory. Giving expectation values to these scalars will deform the
bundle and ``Higgs'' part of the gauge group away. A generic point in
the moduli space will have no gauge group whatsoever.

The analysis of \cite{DMW:hh} indicated that for any such heterotic
string one can find a dual partner. The dual partner is physically the same as
the original but the field identified as the dilaton is
different --- a strongly
coupled string maps to a weakly coupled partner --- and the other
moduli determining the shape of the K3 surface and the moduli of the
bundle undergo a nontrivial transformation. Phrased differently, the
complete moduli space of these heterotic strings, including the
dilaton, admits a nontrivial $\Z_2$ isometry which should be modded
out by in order to obtain the true moduli space of inequivalent
theories in six dimensions.

Now let us compactify this heterotic string on $T^2$. The resulting
theory then has a $U(1)^4$ gauge group --- two $U(1)$'s from the
metric from the isometries of the torus and two from reduction of the
$B$-field. This is one of the examples studied in \cite{KV:N=2}. One
of the $U(1)$'s comes from the $N=2$ supergravity multiplet in four
dimensions and the other three come from vector multiplets. The vector
bundle $E$ described above lives in a moduli space with 448 real
dimensions. Thus, together with the 80 real dimensions of the moduli
space of strings on a K3 surface, our original theory in six
dimensions (without the dilaton) lived in a moduli space of
$448+448+80=976$ dimensions. This gives the moduli space coming from
the hypermultiplets in the four-dimensional theory. This space has a
quaternionic structure --- there are $976/4=244$ hypermultiplets.

For this four dimensional theory to be dual to a type IIA string
compactified on a \CY\ manifold $X$ we therefore require
$h^{1,1}(X)=3$ and $h^{2,1}(X)=243$ (the dilaton of the type II string
gives one hypermultiplet). In \cite{KV:N=2} it was conjectured that
$X$ is given by the blow-up of a hypersurface in a weighted projective
space --- namely the hypersurface given by
\begin{equation}
  x_1^{24}+x_2^{24}+x_3^{12}+x_4^3+x_5^2=0,   \label{eq:d24}
\end{equation}
in the space $\P^4_{\{1,1,2,8,12\}}$.

When this is blown-up, it is a K3 fibration as discussed in
\cite{KLM:K3f}. This then satisfies the condition from \cite{AL:ubiq}
that it be dual to a heterotic string. From the result of
\cite{DMW:hh} however, this is not enough --- we need it to be dual to
two heterotic strings. Thus there must be two K3 fibrations of $X$.
At first sight we are in trouble --- the blow-up of the hypersurface
defined by (\ref{eq:d24}) does {\em not\/} admit two fibrations.

All is not lost however. The manifold $X$ has 243 deformations of
complex structure. The equation (\ref{eq:d24}) only has 242
deformations. Therefore there is one deformation of $X$ which takes it
outside the class of those embedded in $\P^4_{\{1,1,2,8,12\}}$. When
we deform outside this class, the fibration structure of $X$
changes.\footnote{
The best way to see this is as follows: The projection
$\P^4_{\{1,1,2,8,12\}}\rightarrow \P^2_{\{1,1,2\}}$ onto the
first three coordinates induces an elliptic fibration of $X$ onto
$\P^2_{\{1,1,2\}}$. The space $\P^2_{\{1,1,2\}}$ is singular and
should be blown-up to give the ``Hirzebruch surface'' $F_2$.
This elliptic fibration is easily seen to have
a section, and hence is birationally equivalent to a Weierstrass
model
$$y^2=x^3+ax+b,$$
where $a\in \Gamma(\omega_{F_2}^{-4})$ and $b\in\Gamma(\omega_{F_2}^{-6})$.
(See \cite{Nak:We} and also \cite{Gross:ft}.) Now we can deform $F_2$
to $\P^1\times\P^1$, and deform the above Weierstrass model to
$$y^2=x^3+ax+b,$$
where $a\in \Gamma(
\omega_{\P^1\times\P^1}^{-4})$ and $b\in\Gamma(\omega_{\P^1\times\P^1}^{-6})$.
This yields the additional deformation of complex structure not visible in
$\P^4_{\{1,1,2,8,12\}}$.}

The result is that, at a generic complex structure, $X$ is not only a
K3 fibration but can also be
written as an elliptic fibration with base space
$\P^1\times\P^1$. Fix a point in the first $\P^1$ factor. Then the
fibration over the second $\P^1$ defines a K3 surface as an elliptic
fibration. Thus the first $\P^1$ can be seen as the base for writing
the whole threefold as a K3 fibration.
Clearly we could have chosen instead to fix a point first in the
second $\P^1$ factor so $X$ admits two K3 fibrations.

We believe that the fact that there is only one K3 fibration for
special values of the complex structure is probably not very significant
from the point of view of string theory. It appears to be tied up with
the phenomenon observed by Wilson \cite{Wil:Kc} that the K\"ahler cone
for a \CY\ manifold can ``jump'' as one moves about the moduli space
of complex structures as rational curves suddenly jump into
existence. One may worry that this may have adverse effects on
rational curve counts along the lines of \cite{CDGP:} but it turns
out that any such rational curve will give a zero contribution when
the analysis of \cite{AM:rat} is performed. Similarly we expect the
dual heterotic string to be unaffected.

Let us clarify the interpretation of our construction of the fibration
in terms of the
heterotic string. Let us again fix a generic point in the first $\P^1$
and consider building a K3 surface as an elliptic fibration over the
second $\P^1$.
This elliptic fibration has one section. This section thus
defines a rational curve embedded inside this K3 surface. The size of
this rational curve is given by the size of second $\P^1$. Now we fibre
the threefold as a K3 fibration over the first $\P^1$. The result is
shown in figure
\ref{fig:fib1}. Thus the size of one of the $\P^1$'s in the
base $\P^1\times\P^1$ gives the size of the rational curve in the
generic fibre and the other gives the size of the base of the
K3 fibration. As explained in \cite{AL:ubiq} the size of the base
gives the value of the dilaton for the heterotic string. What is the
size of the other $\P^1$ in terms of the heterotic string?

\iffigs
\begin{figure}
  \centerline{\epsfxsize=10cm\epsfbox{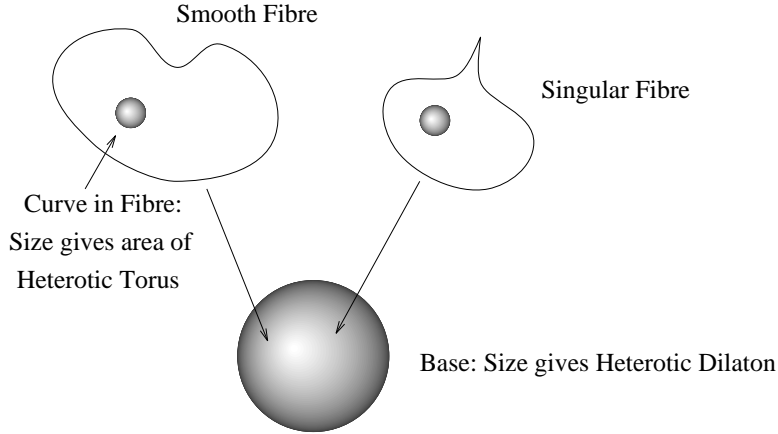}}
  \caption{A K3 fibration with $h^{1,1}=3$.}
  \label{fig:fib1}
\end{figure}
\fi

Let $H_1$ and $H_2$ be our two heterotic strings. In six dimensions
let their string couplings (exponential of the dilaton) be given
by $\lambda_{6,H_1}$ and $\lambda_{6,H_2}$. In \cite{DMW:hh} it was
shown that
\begin{equation}
  \lambda_{6,H_1} = \frac1{\lambda_{6,H_2}}.
\end{equation}
Now let us compactify further down to four dimensions to give a dual
pair of heterotic strings with coupling constants $\lambda_{4,H_1}$
and $\lambda_{4,H_2}$. To do this we compactify on a two-torus. The
four-dimensional coupling depends on the six-dimensional coupling and
the area of this two torus. This area differs according to which
heterotic string measures it \cite{DMW:hh}. Let us denote the area of the torus
$T_1$ or $T_2$ depending on whether it is measured by $H_1$ or $H_2$.
We then have
\begin{equation}
\eqalign{
  \lambda_{4,H_1}^2 &= \frac{\lambda_{6,H_1}^2}{T_1}
   = \frac{\lambda_{6,H_1}^2}{T_2.\lambda_{6,H_1}^2} = \frac1{T_2}\cr
  \lambda_{4,H_2}^2 &= \frac1{T_1}\cr}
\end{equation}
Thus the size of the other $\P^1$ in the type IIA picture is
immediately apparent --- it is the size of the torus in the heterotic
string.

Note that the act of exchanging the fibrations with $X$ generically
leads to a change in the complex structure of $X$. That is, there is a
$\Z_2$ symmetry in the moduli space of complex structures {\em
together\/} with K\"ahler forms but not on either alone. Translated
into the heterotic string language this tells us that as we map from
one heterotic string to its dual partner, there is a non-trivial map
on the moduli space of hypermultiplets. This is exactly what was found
in \cite{DMW:hh}.

For some values of the moduli of the hypermultiplets the exchange of
the base $\P^1$'s will be a symmetry of $X$.
Thus we have arrived at a manifest geometric realization of the
duality between the two heterotic strings in terms of a symmetry of
the space on which the equivalent type IIA string is
compactified. Note that some evidence of the symmetry in the context
of type IIB strings has previously been noted in \cite{KLM:K3f}.

It is interesting to note that we can now go some way to prove the
conjecture made in \cite{KV:N=2} that a type IIA string compactified
on $X$ is indeed dual to the heterotic string above. This is because
one may show that $X$ is the {\em only\/} \CY\ manifold which is an
elliptic fibration over $\P^1\times\P^1$ which has the right Hodge
numbers. To more precise, suppose the following statements hold.
\begin{enumerate}
\item The heterotic-heterotic duality in six dimensions proposed in
\cite{DMW:hh} is correct.
\item This heterotic string compactified on a torus down to four
dimensions is dual to a type IIA string compactified on some manifold
$X_1$.
\item The weakly-coupled heterotic string can be understood in terms
of $X_1$ near its large radius limit.
\end{enumerate}
We can then state that {\em $X_1$ is isomorphic to $X$}.

\def\Pone{\P^1}
Finally let us prove this uniqueness. This is a little technical and
the reader may wish to skip to the next section.
Suppose $Y$ is any Calabi-Yau manifold with two K3 fibrations
and $h^{2,1}(Y)=243$. Then we claim that $Y$ is of the same type
as the $X$ constructed above. Indeed, the two distinct maps
$Y\rightarrow\Pone$ induces an elliptic fibration $f:Y\rightarrow
\Pone\times\Pone$. If $f$ does not have a section, then there is an
elliptic \CY\ threefold $j:J(Y)\rightarrow\Pone\times\Pone$, the Jacobian
of $Y$, which is fiber-wise isomorphic to $f$ except for possibly
some isolated fibers, and such that $j$ has a section. See \cite{Gross:ft}
for details. Then $j$ is birationally equivalent to a Weierstrass model
over $\P^1\times\P^1$ as in the footnote above. Since
$h^{1,2}(Y)=h^{1,2}(X)$ and
each $J(Y)$ can arise from only a finite number of different
$Y'$ in the same deformation class of $Y$ (as follows from the
theory of \cite{Gross:ft}), we see that we obtain a dominant map from
the moduli space of $Y$ to the moduli space of $X$. However, again using
the theory of \cite{Gross:ft}, for general $X$, there is no elliptic
fibration $Y\rightarrow\Pone\times\Pone$ without a section whose
Jacobian is $X$. Thus $j:Y\rightarrow\P^1\times\P^1$ must already have
a section, so $Y$ is deformation equivalent to $X$.


\section{Enhanced Gauge Groups}		\label{s:enhg}

It was discussed in \cite{DMW:hh} that going between the dual
heterotic strings exchanges the r\^oles of the perturbative gauge
group and the nonperturbative gauge group. We will now see that this
is also in accord with the double fibration structure.

The construction of the heterotic string in question began with a
theory with gauge group $E_7\times E_7$. Thus was then removed by
picking a generic value for the bundle moduli. Thus we know that the
$U(1)^4$ gauge group in our four-dimensional theory can be enhanced to
at least $E_7\times E_7\times U(1)^4$ by tuning the hypermultiplet moduli
suitably. When we are at such a point of enhanced symmetry, some
vector moduli become massless and we can move into a new branch of the
moduli space roughly along the lines of \cite{GMS:con}. The new theory
at a generic point in its moduli space will have rank$(E_7\times E_7)=14$
more vector multiplets than the original and fewer hypermultiplets
since we had to specialize a bundle with structure group $E_8\times
E_8$ to one of $SU(2)\times SU(2)$. An $SU(2)$ bundle with $c_2=12$
has a moduli space of 84 real dimensions and so the number of
hypermultiplets lost is $(448-84)*2/4=182$. Thus if this new heterotic
string is dual to a type IIA string on a \CY\ manifold $X^\prime$, we
have $h^{1,1}(X^\prime) = 17$ and $h^{2,1}(X^\prime) = 61$.

The appearance of enhanced gauge symmetries in the context of type II
string compactifications has been discussed in
\cite{BSV:D-man,me:en3g,KMP:enhg}.
As seen from this new heterotic string, the $E_7\times E_7$ gauge
group appears by varying the vector moduli. For our purposes it will
be the analysis of \cite{me:en3g} which will be of most use. There it
was claimed that the enhanced
gauge groups which could be seen perturbatively by the heterotic
string as the vector moduli varied must originate from singularities
in the K3 fibre. Thus, the $E_7\times E_7$ gauge group appears when
the generic K3 fibre has two $E_7$ singularities in it
\cite{W:dyn}. So long as this extremal transition between
$X$ and $X^\prime$ preserves the K3 fibration structure then we can
see that the $E_7\times E_7$ gauge group appears in the original
heterotic string when we deform the {\em complex structure\/} of $X$
so as to acquire two $E_7$ singularities in the generic fibre.

We can do this explicitly as follows. Let us consider $X$ as an elliptic
fibration over $\P^1\times\P^1$. When viewed as a K3 fibration, the
generic fibre will itself be an elliptic fibration. Let us denote the
homogeneous coordinates of the base $\P^1\times\P^1$ by $[s_0,s_1]$ and
$[t_0,t_1]$. A fibre can then be written in Weierstrass form as
\begin{equation}
  y^2 = x^3+a(s_i,t_i)x+b(s_i,t_i).	\label{eq:efib}
\end{equation}
In order for the resulting threefold to be a \CY\ manifold we require
that $a$ is of bidegree $(8,8)$ in $s_i$ and $t_i$ and that $b$ be of
bidegree $(12,12)$ \cite{Gross:ft}.

As we move over the base $\P^1\times\P^1$, the discriminant of the
elliptic, $4a^3+27b^2$, will have zeros leading to singular fibres. If
the polynomials $a$ and $b$ are generic, the singular fibres will not
be too bad and the resulting \CY\ manifold will be smooth. For special
choices of the polynomials however, the singular fibre will be
sufficiently bad as to introduce singularities into the \CY\
threefold. This has been discussed in the context of string theory and
$D$-branes in \cite{BSV:D-man}.

\iffigs
\begin{figure}
  \centerline{\epsfxsize=10cm\epsfbox{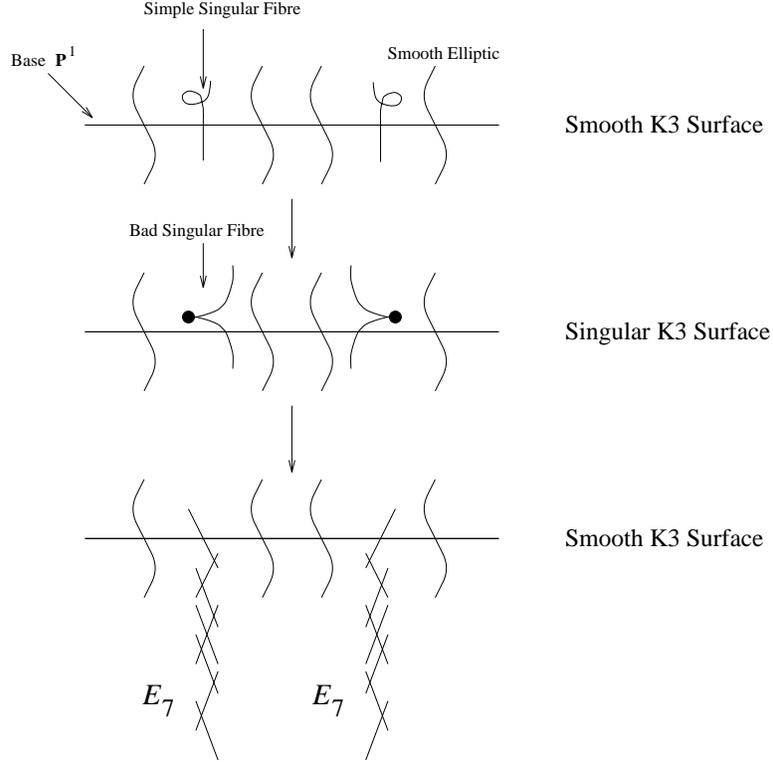}}
  \caption{A transition of the generic K3 fibre.}
  \label{fig:K3trans}
\end{figure}
\fi

Let us fix a generic point on one of the $\P^1$'s (i.e., specify $t_0$
and $t_1$) in the base. The
elliptic fibration given in (\ref{eq:efib}) will then give the generic
K3 fibre as an elliptic fibration. We now want to deform this as to
get two $E_7$ singularities in the K3 surface. Following the work of
\cite{Mir:fibr} we can do this by choosing the following polynomials
for $a$ and $b$:
\begin{equation}
\eqalign{
  a &= s_0^3s_1^3 f(s_i,t_i)\cr
  b &= s_0^5s_1^5 g(s_i,t_i),\cr}
\end{equation}
where $f$ is of bidegree $(2,8)$ and $g$ is of bidegree $(2,12)$. We
now claim that the generic K3 fibre has an $E_7$ singularity at $s_0=0$
and another at $s_1=0$. We can now blow each K3 fibre up to resolve
this singularity.

This process of deforming the complex structure to acquire a
singularity and then blowing it up is shown for a generic K3 fibre in
figure \ref{fig:K3trans}. We show complex dimensions as real. In the
each case the K3 surface is an elliptic fibration over the base
$\P^1$. The blowing up procedure introduces two sets of rational
curves in the shape of the Dynkin diagram for $E_7$.

In terms of the generic K3 fibre, this process has simply changed one
K3 surface into another. The effect globally on the \CY\ threefold
is not so trivial however. We need to worry that the
degenerate fibres of the K3 fibration may be sufficiently bad as to
introduce singularities into the \CY\ threefold. It may be shown that
this is not the case here. Our new smooth \CY\ manifold obtained by
this extremal transition is, of course, the obvious candidate for
$X^\prime$. Indeed one can show that it satisfies $h^{1,1}=17$ and
$h^{2,1}=61$.

We claim then that we have constructed the \CY\ manifold $X^\prime$
which is dual to the heterotic string obtained from the $E_7\times
E_7$ heterotic string deformed by vector moduli to obtain a string with
gauge group $U(1)^{4+14}$. Thus we have another example to add to those in
\cite{KV:N=2,me:flower} of the geometric interpretation of a phase
transition in the heterotic string in terms of the type IIA string.

Now let us ask what happens when we exchange the $\P^1$'s in the base
of the elliptic fibration as we try to go through this extremal
transition. The base of the K3 fibration is now what was the base of
generic K3 fibre when it was written as an elliptic fibration. Thus,
we have our $E_7$ trees sticking out over a finite number of points on
this base as shown in the bottom of figure \ref{fig:K3trans}. That is,
this contribution to the counting of the vector
multiplets comes from having reducible bad fibres in the K3
fibration. This is exactly the situation where we expect to have
nonperturbative contributions to the gauge group of the heterotic
string as explained in \cite{AL:ubiq}. Thus we see that the $E_7\times
E_7$ that was perturbative in the original heterotic string becomes
nonperturbative in the dual heterotic string. What's more, one can
also see that the generic fibre of the second fibration will not
contain any singularities and so the perturbative gauge group consists
only of the $U(1)^4$ which is always present.

It is easy to extend this analysis to the simply-laced gauge groups which are
subgroups within $E_7\times E_7$ and in each case we see the structure
conjectured in \cite{DMW:hh} appearing. That is, the parts of the
gauge group which are perceived as perturbative and nonperturbative
are exchanged (except for the $U(1)^4$) between the dual heterotic strings.

Up to this point we have managed to reproduce the effects seen in
\cite{DMW:hh} in terms of the geometry of the type IIA string. It
turns out that we can say more however. The same method as was used to
introduce two $E_7$ singularities into the generic K3 fibre can
actually be used to introduce two $E_8$ singularities. Thus we appear
to be claiming that the heterotic string can be deformed to
a theory with a gauge group containing $E_8\times E_8$. This is troublesome to
interpret in terms of the usual picture since the centre of $E_8\times
E_8$ is trivial --- we need to deform our $E_8\times E_8$ bundle so
that it becomes trivial and yet maintain $c_2=24$. Clearly this is
impossible. What we can do however in addition is to bend the base K3
around and possibly introduce singularities into the base, or the
bundle (perhaps reinterpreting it as a sheaf). (One may also try to
analyze this in terms of open strings along the lines of \cite{GP:open}.)
Such processes can
introduce further gauge groups. One should also suspect that this may
introduce nonperturbative gauge groups however following Witten's work
\cite{W:small-i}. We will now see that this is indeed the case.

Something new happens when we degenerate to an $E_8\times E_8$
singularity compared to this cases of $E_7\times E_7$ and its
subgroups. That is, when we blow-up to resolve the generic fibre it
turns out that some singular fibres are still sufficiently badly-behaved
as to make the resulting \CY\ threefold singular. If we want to obtain
a smooth threefold we require further blow-ups within these bad fibres. Thus
we necessarily have contributions to $h^{1,1}$ of the threefold from
bad fibres and thus, following \cite{AL:ubiq}, we have a
nonperturbative part to the gauge group.

In the case at hand one can show that each of six bad fibres for each
$E_8$ need to
be blown up once to smooth the threefold. This suggests that we have a
nonperturbative gauge group of rank 12 when we try to enhance the
perturbative part of the gauge group to include $E_8\times E_8$. It is nice to
see that a nonperturbative contribution to the gauge group arises when
one tries to enhance to $E_8\times E_8$ but we have no definite
geometric interpretation for this on the heterotic side yet in terms
of singular bundles and/or singular K3 surfaces. We should say that
finding the heterotic string description of this model is necessarily
going to be difficult since there is always some nonperturbative
contribution to the gauge group for any point in its moduli space.


\section{An Extreme Case}		\label{s:inf}

We have discussed above an example of a \CY\ manifold which admits two
K3 fibrations. Actually there is no limit to the number of K3
fibrations one may obtain and in this section we discuss an example
with an infinite number. In a way this will turn out to be more
trivial than the previous case.

When one has more than one K3 fibration, the relationship between the
different fibrations can be of one of three types:
\begin{enumerate}
\item The fibrations have quite different generic fibres. That is, the
Picard lattice for the generic fibres differ.
\item The generic fibres have the same Picard lattice but, for a generic
complex structure, the act of exchanging the fibrations has a
nontrivial action on the complex structure.
\item The fibrations are completely diffeomorphic.
\end{enumerate}

An example of the first case is $X^\prime$ from section
\ref{s:enhg}. After going through the $E_7\times E_7$ transition, the
Picard lattice the generic K3 fibre will now contain the root lattice
of $E_7\times E_7$ whereas the generic fibre of the other fibre will
not. In terms of the heterotic string case 1 corresponds to the
perturbative part of the gauge group for one heterotic string being
different to the
perturbative part for the other. In the case of $X^\prime$, one string
has a large perturbative gauge group of rank 18, whereas its dual
partner only sees $U(1)^4$.

An example of the second case is $X$. Both heterotic strings have a
perturbative gauge group of $U(1)^4$ but, as discussed in
\cite{DMW:hh}, the map between them is nontrivial on the
hypermultiplet moduli space.

We will now discuss an example of the third type following
\cite{Og:K3f}. Consider the \CY\
manifold, $Y$, obtained by blowing up the orbifold
$T^6/(\Z_2\times\Z_2)$. This orbifold has been discussed many times in
the string literature, for example \cite{VW:tor}.

The first $\Z_2$ quotient may be thought of as building a K3 surface
as $T^4/\Z_2$. The second quotient then acts on this K3 surface and
the other $T^2$ to build a \CY\ space. Let us analyze this
intermediate K3 surface and in particular its Picard lattice,
$\Gamma$ (see, for example \cite{BPV:} for a discussion of the Picard
lattice in this context). The two $T^2$'s used to build this K3 surface each
contribute a rational curve class to the K3 surface as well as the 16
exceptional divisors coming from blowing up the orbifold. Thus, for a
generic model, $\Gamma$ is a rank 18 lattice of signature
$(1,17)$. Note that this lattice is {\em not\/} self-dual.

Now, thanks to the work of \cite{PSS:tor}, it turns out that any primitive
element of length 0 in $\Gamma$ defines the class of an elliptic curve
in the K3
surface and that there is an elliptic fibration of the K3 surface
whose generic fibre is this class. Thus, since there are an infinite
number of such vectors, there are an infinite number of elliptic
fibrations for such a K3 surface.

Now divide out by the second $\Z_2$. There are two possibilities for
how this $\Z_2$ acts on our elliptic fibration of the K3 surface. It
may either fix the base $\P^1$ point-wise or it may not. In the case
that the base $\P^1$ is fixed, one may show that this may be used as
the base of $Y$ as a K3 fibration \cite{Og:K3f}. Loosely speaking,
this happens for ``half'' of the primitive elements of length zero in
$\Gamma$. Thus $Y$ has an infinite number of K3 fibrations.

If our assertion in the introduction concerning the existence of a
heterotic string for every K3 fibration is correct then we have
an infinite number of heterotic strings any two of which are dual to
each other!

Actually this is not as bad as it sounds. One can see that a subgroup
(of finite index) of $O(1,17;\Z)$ must act as a symmetry group on this
\CY\ manifold since it acted as such on the K3 surface above. This
can be used to identify all the different K3 fibrations we have
generated. Thus all the K3 fibrations are diffeomorphic.

One should note that this does not mean that this example is trivial
however. One should think of this subgroup of $O(1,17;\Z)$ as acting
as a group of $U$-dualities on the heterotic string in the sense of
\cite{HT:unity}. The group acts nontrivially on the dilaton and so
identifies heterotic strings that do not look isomorphic from their
conformal field theories. One should also note that it mixes
contributions to $h^{1,1}$ from generic fibres and from bad fibres and
so mixes perturbative gauge groups with nonperturbative gauge groups
as in the earlier example. The difference between this example and
that in section \ref{s:eg1} is that this time the duality maps acts
trivially on the space of hypermultiplets.

In \cite{Mor:cp} a conjecture was made about \CY\ manifolds with such
infinite symmetries. If it is correct then it will always be the case that
a \CY\ manifold with an infinite number of fibrations will lead to
heterotic strings which are mainly equivalent in this manner. Dividing
out by this group of dualities one can be left with only a finite
number of heterotic strings which are equivalent in the sense of
type 1 or type 2 above.


\section*{Acknowledgements}

We would like to thank M.~Duff, S.~Kachru, V.~Kaplunovsky and
D.~Morrison for useful conversations.
The work of the authors is supported by grants from
the National Science Foundation.


\begin{thebibliography}{10}

\bibitem{HT:unity}
C.~Hull and P.~Townsend,
\newblock {\em Unity of Superstring Dualities},
\newblock Nucl. Phys. {\bf B438} (1995) 109--137.

\bibitem{W:dyn}
E.~Witten,
\newblock {\em String Theory Dynamics in Various Dimensions},
\newblock Nucl. Phys. {\bf B443} (1995) 85--126.

\bibitem{KV:N=2}
S.~Kachru and C.~Vafa,
\newblock {\em Exact Results For N=2 Compactifications of Heterotic Strings},
\newblock Nucl. Phys. {\bf B450} (1995) 69--89.

\bibitem{FHSV:N=2}
S.~Ferrara, J.~Harvey, A.~Strominger, and C.~Vafa,
\newblock {\em Second Quantized Mirror Symmetry},
\newblock Phys. Lett. {\bf 361B} (1995) 59--65.

\bibitem{KLM:K3f}
A.Klemm, W.Lerche, and P.Mayr,
\newblock {\em K3--Fibrations and Heterotic-Type II String Duality},
\newblock Phys. Lett. {\bf 357B} (1995) 313--322.

\bibitem{VW:pairs}
C.~Vafa and E.~Witten,
\newblock {\em Dual String Pairs With $N=1$ and $N=2$ Supersymmetry in Four
  Dimensions},
\newblock Harvard and IAS 1995 preprint HUTP-95/A023, hep-th/9507050.

\bibitem{AL:ubiq}
P.~S. Aspinwall and J.~Louis,
\newblock {\em On the Ubiquity of K3 Fibrations in String Duality},
\newblock Cornell and Munich 1995 preprint CLNS-95/1369, LMU-TPW 95-16,
  hep-th/9510234,
\newblock to appear in Phys. Lett. {\bf B}.

\bibitem{Mor:gid}
D.~R. Morrison,
\newblock {\em Mirror Symmetry and Rational Curves on Quintic Threefolds: A
  Guide For Mathematicians},
\newblock J. Amer. Math. Soc. {\bf 6} (1993) 223--247.

\bibitem{DMW:hh}
M.~J. Duff, R.~Minasian, and E.~Witten,
\newblock {\em Evidence for Heterotic/Heterotic Duality},
\newblock Texas A\&M 1996 preprint CTP-TAMU-54/95, hep-th/9601036.

\bibitem{MV:F}
D.~R. Morrison and C.~Vafa,
\newblock {\em Compactifications of F-Theory on Calabi--Yau Threefolds --- I},
\newblock to appear.

\bibitem{Nak:We}
N.~Nakayama,
\newblock {\em On Weierstrass Models},
\newblock in ``Algebraic Geometry and Commutative Algebra in Honor of Masayoshi
  Nagata'', pages 405--431, Kinokuniya, Tokyo, 1988.

\bibitem{Gross:ft}
M.~Gross,
\newblock {\em A Finiteness Theorem for Elliptic Calabi--Yau Threefolds},
\newblock Duke Math. J. {\bf 74} (1994) 271--299.

\bibitem{Wil:Kc}
P.~M.~H. Wilson,
\newblock {\em The K{\"a}hler Cone on Calabi--Yau Threefolds},
\newblock Invent. Math. {\bf 107} (1992) 561--583.

\bibitem{CDGP:}
P.~Candelas, X.~C. de~la Ossa, P.~S. Green, and L.~Parkes,
\newblock {\em A Pair of Calabi--Yau Manifolds as an Exactly Soluble
  Superconformal Theory},
\newblock Nucl. Phys. {\bf B359} (1991) 21--74.

\bibitem{AM:rat}
P.~S. Aspinwall and D.~R. Morrison,
\newblock {\em Topological Field Theory and Rational Curves},
\newblock Commun. Math. Phys. {\bf 151} (1993) 245--262.

\bibitem{GMS:con}
B.~R. Greene, D.~R. Morrison, and A.~Strominger,
\newblock {\em Black Hole Condensation and the Unification of String Vacua},
\newblock Nucl. Phys. {\bf B451} (1995) 109--120.

\bibitem{BSV:D-man}
M.~Bershadsky, V.~Sadov, and C.~Vafa,
\newblock {\em D-Strings and D-Manifolds},
\newblock Harvard 1995 preprint HUTP-95/A035, hep-th/9510225.

\bibitem{me:en3g}
P.~S. Aspinwall,
\newblock {\em Enhanced Gauge Symmetries and Calabi-Yau Threefolds},
\newblock Cornell 1995 preprint CLNS-95-1375, hep-th/9511171,
\newblock to appear in Phys. Lett. {\bf B}.

\bibitem{KMP:enhg}
S.~Katz, D.~R. Morrison, and M.~R. Plesser,
\newblock {\em Enhanced Gauge Symmetry in Type II String Theory},
\newblock Oklahoma et al 1996 preprint OSU-M-96-1, hep-th/9601108.

\bibitem{Mir:fibr}
R.~Miranda,
\newblock {\em Smooth Models for Elliptic Threefolds},
\newblock in R.~Friedman and D.~R. Morrison, editors, ``The Birational Geometry
  of Degenerations'', Birkhauser, 1983.

\bibitem{me:flower}
P.~S. Aspinwall,
\newblock {\em An $N=2$ Dual Pair and a Phase Transition},
\newblock Cornell 1995 preprint CLNS-95/1366, hep-th/9510142,
\newblock to appear in Nucl. Phys. {\bf B}.

\bibitem{GP:open}
E.~G. Gimon and J.~Polchinski,
\newblock {\em Consistency Conditions for Orientifolds and D-Manifolds},
\newblock ITP 1996 preprint NSF-ITP-96-01, hep-th/9601038.

\bibitem{W:small-i}
E.~Witten,
\newblock {\em Small Instantons in String Theory},
\newblock IAS 1995 preprint IASSNS-HEP-95-87, hep-th/9511030.

\bibitem{Og:K3f}
K.~Oguiso,
\newblock {\em On Algebraic Fiber Space Structures on a Calabi-Yau 3-fold},
\newblock Int. J. of Math. {\bf 4} (1993) 439--465.

\bibitem{VW:tor}
C.~Vafa and E.~Witten,
\newblock {\em On Orbifolds with Discrete Torsion},
\newblock J. Geom. Phys. {\bf 15} (1995) 189--214.

\bibitem{BPV:}
W.~Barth, C.~Peters, and A.~van~de Ven,
\newblock {\em Compact Complex Surfaces},
\newblock Springer, 1984.

\bibitem{PSS:tor}
I.~I. Piatecki{\u\i}-Shapiro and I.~R. Shafarevich,
\newblock {\em A Torelli Theorem for Algebraic Surfaces of Type K3},
\newblock Math. USSR Izv. {\bf 5} (1971) 547--588.

\bibitem{Mor:cp}
D.~R. Morrison,
\newblock {\em Compactifications of Moduli Spaces Inspired by Mirror Symmetry},
\newblock in A.~Beauville et~al., editors, ``Journ\'ees de G\'eom\'etrie
  Alg\'ebrique d'Orsay (Juillet 1992)'', volume 218 of Ast\'erisque, pages
  243--271, Soci\'et\'e Math\'ematique de France, 1993.

\end{thebibliography}

\end{document}